\documentclass[multphys,vecarrow]{svmult}

\usepackage{graphicx}
\usepackage{amssymb}
\usepackage{amsmath}
\usepackage{subeqnar}
\usepackage{epsfig}
\usepackage{psfrag}
\usepackage{citesort}

\begin{document}

\frontmatter

\mainmatter

\title*{
  A Comparison of a Cellular Automaton and a Macroscopic Model
}

\author{
  Sven Maerivoet\inst{1}, Steven Logghe\inst{2},
  Bart De Moor\inst{1}, and Ben Immers\inst{3}
}

\institute{	
  Katholieke Universiteit Leuven\\
  Department of Electrical Engineering ESAT-SCD (SISTA)\\
  Kasteelpark Arenberg 10, B-3001 Leuven, Belgium\\
  Phone: +32 (0) 16 32 19 70\\
  E-mail: {\tt sven.maerivoet@esat.kuleuven.ac.be}\\
  \and
  Transport \& Mobility Leuven\\
  Tervuursevest 54 bus 4, B-3000 Leuven, Belgium\\
  Phone: +32 (0) 16 22 95 52\\
  E-mail: {\tt steven.logghe@tmleuven.be}\\
  \and
  Katholieke Universiteit Leuven\\
  Department of Civil Engineering (Transportation Planning and Highway Engineering)\\
  Kasteelpark Arenberg 40, B-3001 Leuven, Belgium\\
  Phone: +32 (0) 16 32 16 73\\
}

\authorrunning{
  S. Maerivoet, S. Logghe, B. De Moor, and B. Immers
}

\date{
  \today
}

\maketitle

% ========
% ABSTRACT
% ========
\begin{abstract}
  In this paper we describe a relation between a microscopic stochastic traffic 
  cellular automaton model (i.e., the STCA) and the macroscopic first-order 
  continuum model (i.e., the LWR model). The innovative aspect is that we 
  explicitly incorporate the STCA's stochasticity in the construction of the 
  fundamental diagram used by the LWR model. We apply our methodology to a small 
  case study, giving a comparison of both models, based on simulations, 
  numerical, and analytical calculations of their tempo-spatial behavior.
\end{abstract}

\section{Introduction}

Dating back to the mid '50s, Lighthill, Whitham, and Richards introduced their 
macroscopic first-order continuum model (i.e., the LWR model) 
\cite{LIGHTHILL:55,RICHARDS:56}. It is based on a fluid dynamics analogy, in 
which the collective behaviour of infinitesimally small particles is described, 
using aggregate quantities such as flow $q$, density $k$ and (space) mean speed 
$\overline v_{s}$. Models like this, can be solved using cell-based numerical 
schemes (e.g., using the Godunov scheme \cite{DAGANZO:95,LEBACQUE:96}).

Later, microscopic traffic flow models have been developed that explicitly 
describe vehicle interactions at a high level of detail. During the early 
nineties, these models were reconsidered from an angle of particle physics: 
cellular automata models were applied to traffic flow theory, resulting in fast 
and efficient modelling techniques for microscopic traffic flow models 
\cite{NAGEL:92}. These cellular automata models can be looked upon as a particle 
based discretisation scheme for macroscopic traffic flow models.

It is from this latter point of view that our paper addresses the common 
structure between the seminal STCA model and the first-order LWR model. Our main 
goal is to provide a means for explicitly incorporating the STCA's stochasticity 
in the LWR model.  After explaining the methodology of our approach, we present 
an illustrative case study that allows us to compare the tempo-spatial 
behavioural results obtained with both modelling techniques.

\section{Methodology}

Already, relations between both types of models (i.e., STCA and LWR) have been 
investigated (e.g., \cite{NAGEL:96}). Our approach is however different, in that 
it provides a practical methodology for specifying the fundamental diagram to 
the LWR model. Assuming that a stationarity condition holds on the STCA's rules, 
this allows us to to incorporate the STCA's stochasticity directly into the 
LWR's fundamental diagram.

We assume that we have the ruleset of the STCA available, as well as the maximum 
allowed speed $v_\text{max}$ and the stochastic noise term $p$ (i.e., the 
\emph{slowdown probability}). Furthermore, a discretisation is given, expressed 
by the cell length $\Delta X = 7.5$~m, the time step $\Delta T = 1$~s, and its 
coupled speed increment $\Delta V = \Delta X \div \Delta T = 27$~km/h.

Relating both the STCA and the LWR models is done using a simple two-step 
approach, in which we first rewrite the STCA's rules (assuming a stationarity 
condition holds), and then convert these new rules into a space gap/speed  
diagram (which is equivalent to a stationary density/flow fundamental diagram).

  \subsection{Rewriting the STCA's rules}
  \label{subsec:RewritingSTCARules}

Starting from the ruleset of the STCA, we rewrite it using a min-max 
formulation. Instead of having several individual rules that give a discrete 
speed, we now have one rule that returns a continuous speed (for an individual 
vehicle):

\begin{eqnarray}
\label{eq:MinMaxRule}
  v(t + \Delta T) & = & p \cdot \min{ \lbrace v(t), g_{s}(t) - 1, v_\text{max} - 1 \rbrace } \nonumber\\
                  & + & (1 - p) \cdot \min{ \lbrace v(t) - 1, g_{s}(t), v_\text{max} \rbrace },
\end{eqnarray}

\noindent
with $v(t + \Delta T) = \max{ \lbrace 0, v(t + \Delta T) \rbrace }$. The 
stationarity condition previously mentioned, asserts that the speed $v(t)$ of a 
vehicle at time $t$ is the same as its speed at time $(t + \Delta T)$:

\begin{equation}
  v(t + \Delta T) = v(t).
\end{equation}

This allows us to reformulate equation (\ref{eq:MinMaxRule}) as a set of linear 
inequalities that express constraints on the relations between $v(t)$, 
$v_\text{max}$, $p$ and the space gap $g_{s}(t)$.

  \subsection{Deriving the fundamental diagram}

\begin{figure}[!ht]
  \centering
  \psfrag{speed}[][]{\scriptsize{Speed}}
  \psfrag{flow}[][]{\scriptsize{Flow}}
  \psfrag{space headway}[][]{\scriptsize{Space headway}}
  \psfrag{density}[][]{\scriptsize{Density}}
  \psfrag{deterministic}[][]{\scriptsize{Deterministic}}
  \psfrag{stochastic}[][]{\scriptsize{Stochastic}}
  \psfrag{kc-1}[][]{\scriptsize{$k_{c}^{-1}$}}
  \psfrag{kc}[][]{\scriptsize{$k_{c}$}}
  \psfrag{kj}[][]{\scriptsize{$k_{j}$}}
  \includegraphics[width=12cm,height=4cm]{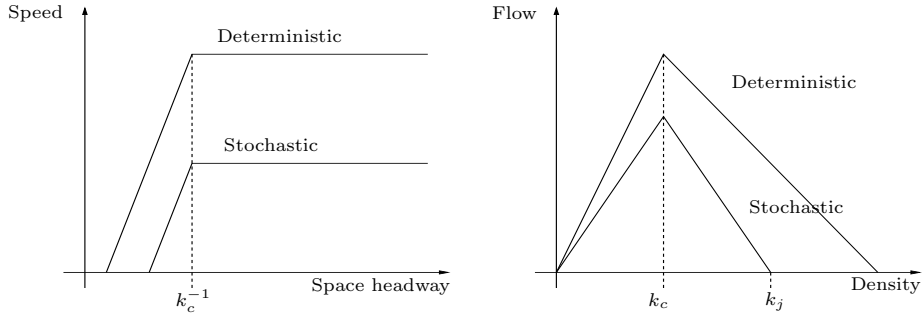}
  \caption{
    Deriving stationary ($h_{s}$,$v$) (left) and ($k$,$q$) (right) fundamental 
    diagrams for the LWR model, after incorporation of the STCA's stochastic 
    noise.
  }
  \label{fig:StationaryHV-KV-FDs}
\end{figure}

The linear inequalities derived in section \ref{subsec:RewritingSTCARules} 
together form a set of boundaries that can be plotted in a diagram that shows 
the space gap $g_{s}$ of a vehicle versus its speed $v$. Knowing that the space 
headway $h_{s}$ equals the vehicle's length $L$ plus its space gap $g_{s}$, we 
can plot a stationary ($h_{s}$,$v$) diagram as can be seen in the left part of 
figure \ref{fig:StationaryHV-KV-FDs}. Because the space headway $h_{s}$ is 
inversely proportional to the density $k$, we can derive an equivalent 
\emph{triangular} ($k$,$q$) fundamental diagram, corresponding to the right part 
of figure \ref{fig:StationaryHV-KV-FDs}.

Considering the previously derived constraints and the diagrams in figure 
\ref{fig:StationaryHV-KV-FDs}, the following important observations can be made:

\begin{itemize}
  \item The stochastic effects from the STCA, are now incorporated in a
        stationary fundamental diagram, which can then be specified as a
        parameter to the LWR model.
  \item The stochastic diagrams lie lower than their deterministic
        counterparts, so the capacity flow is lower for the stochastic variants.
  \item The jam density for stochastic systems is different from that for
        deterministic systems, but the critical density remains unchanged.
\end{itemize}

\section{An illustrative case study}

As a toy example, we apply our methodology to a case study, in which we model a 
single lane road that has a middle part with a reduced maximum speed (e.g., an 
elevation, or a speed limit, \ldots). The road consists of three consecutive 
segments $A$, $B$, and $C$. The first segment $A$ consists of 1500 cells 
(11.25~km), while the second and third segments $B$ and $C$ each consist of 750 
cells (i.e., each approximately 5.6~km long). We consider a time horizon of 
2000~s. The maximum speed for segments $A$ and $C$ is 5~cells/s, whereas it's 
2~cells/s for segment $B$. The stochastic noise $p$ was set to 0.1 for all three 
segments. Vehicles enter the road at segment $A$, travel through segment $B$, 
and exit at the end of segment $C$.

\begin{figure}[!ht]
  \centering
  \frame{\includegraphics[width=8cm]{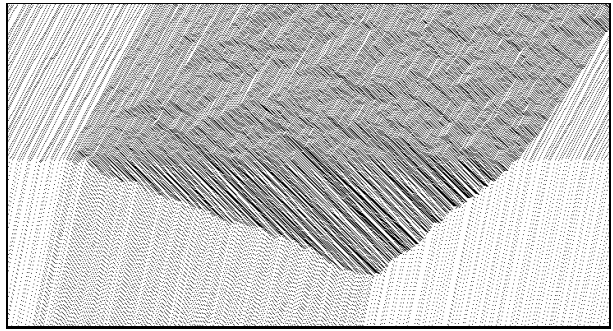}}
  \caption{
    A time/space diagram after simulation of the STCA: each vehicle is 
    represented by a single dot (the time and space axes are oriented 
    horizontally, respectively vertically). At the end of segment $A$, we can 
    see the formation and dissolution of an upstream growing congested region, 
    related to the short traffic burst.
  }
  \label{fig:TCATrajectoriesCloseup}
\end{figure}

This road is simulated using both the STCA and the LWR model. As for the 
boundary conditions, we assume an overall inflow of $q_{c}^{B} \div 2$ ($q_{c}$ 
is the capacity flow), except from $t = 200$~s to $t = 600$~s, where we create a 
short traffic burst with an inflow of $(q_{c}^{A,C} + q_{c}^{B}) \div 2$. Figure 
\ref{fig:TCATrajectoriesCloseup} shows the individual vehicle trajectories in a 
time/space diagram: heavy congestion sets in and flows upstream into segment 
$A$, where it starts to dissolve at the end of the traffic burst.

\begin{figure}[!ht]
  \centering
  {
    \psfrag{Link A}[l][l]{~~\scriptsize{Link A}}
    \psfrag{Link B}[l][l]{~~\scriptsize{Link B}}
    \psfrag{Link C}[l][l]{~~\scriptsize{Link C}}
    \includegraphics[width=5.7cm]{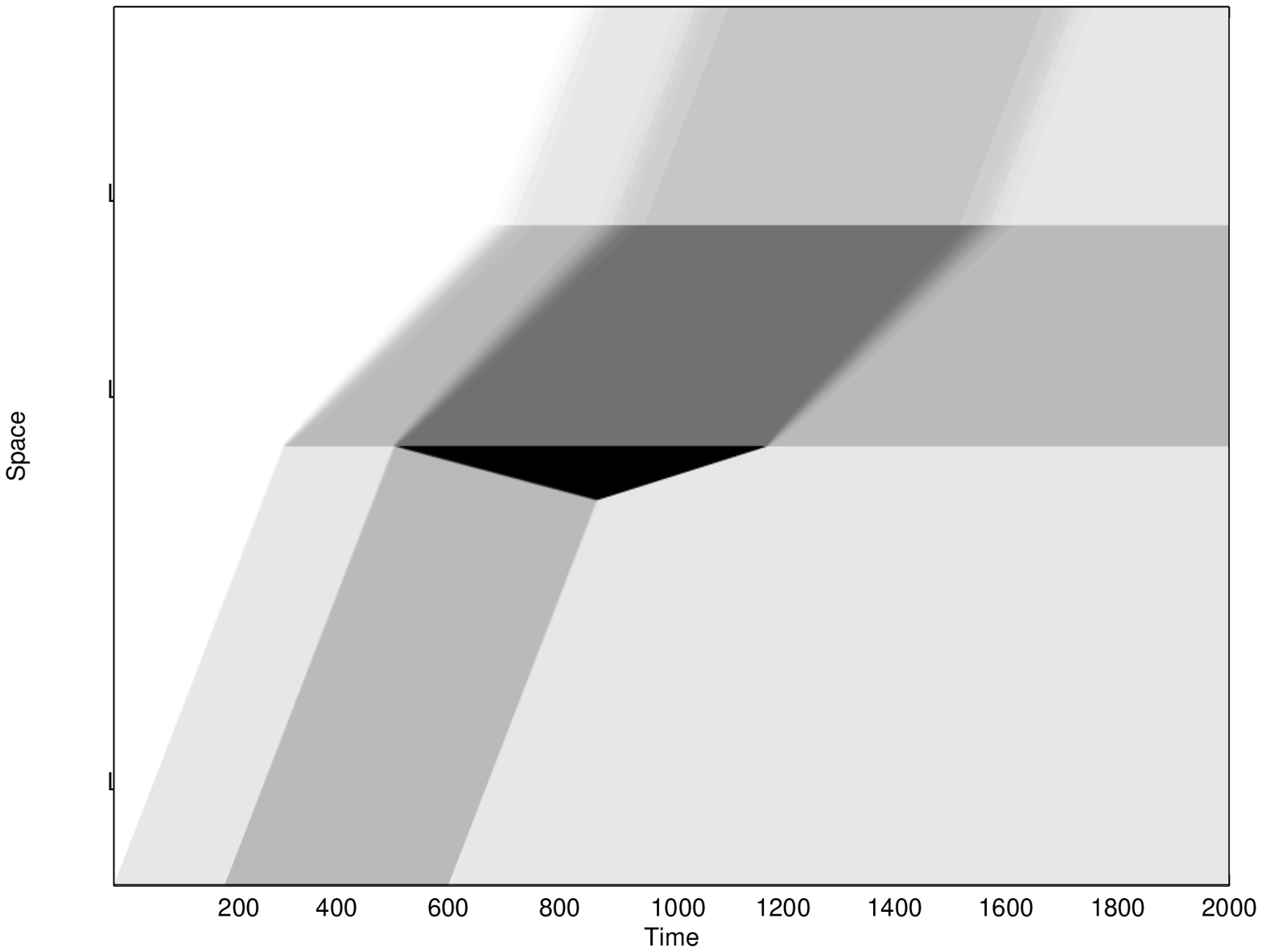}
  }
  {
    \psfrag{Link A}[l][l]{~~~~~\scriptsize{Link A}}
    \psfrag{Link B}[l][l]{~~~~~\scriptsize{Link B}}
    \psfrag{Link C}[l][l]{~~~~~\scriptsize{Link C}}
    \includegraphics[width=5.7cm]{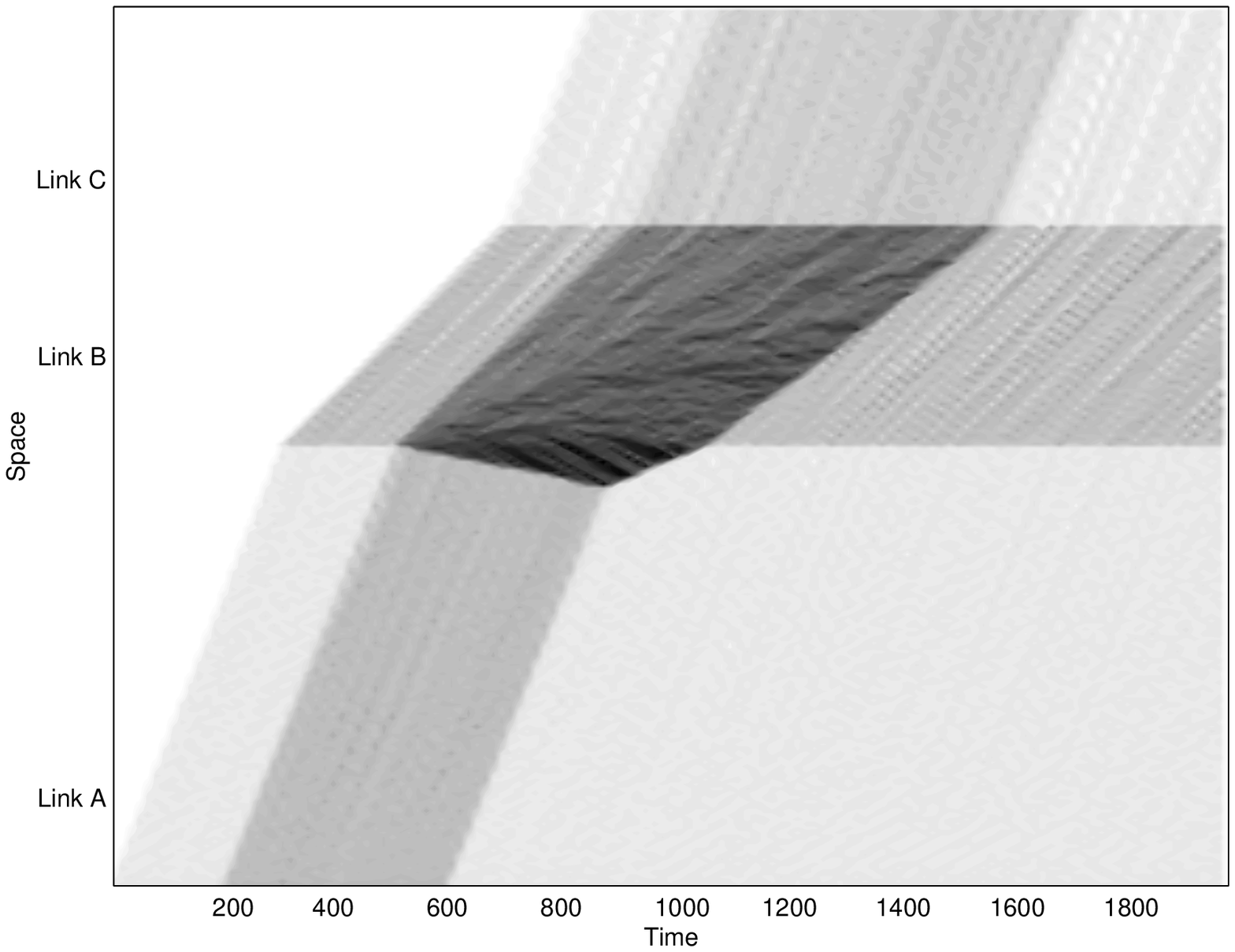}
  }
  \caption{
    Time/space diagrams showing the propagation of densities during 2000~s for 
    the road in the case study. The left part shows the results for the LWR 
    model, while the right part shows the microscopic system dynamics from the 
    STCA model (note that darker regions correspond to congested traffic).
  }
  \label{fig:LWRAndTCATXDensityDiagrams}
\end{figure}

Applying our previously discussed methodology, we construct a stationary 
($k$,$q$) fundamental diagram, and numerically solve the LWR model. The result 
can be seen in the left part of figure \ref{fig:LWRAndTCATXDensityDiagrams}. 
Comparing this spatio-temporal behaviour of the LWR model with the microscopic 
system dynamics from the STCA model (i.e., the right part of figure 
\ref{fig:LWRAndTCATXDensityDiagrams}), we find a qualitatively good agreement 
between the two approaches.

Even more interesting, is the fact that the STCA model reveals a higher-order 
effect that is not visible in the LWR model: there exists a fan of forward 
propagating density waves in segment $B$. Furthermore, in its tempo-spatial 
diagram, the STCA seems to be able to visualise the characteristics that 
constitute the solution of the LWR model.

\section{Conclusions}

The novel approach taken in our research, allows us to incorporate the STCA's 
stochasticity directly in the first-order LWR model. This is accomplished by 
means of a stationarity condition that converts the STCA's rules into a set of 
linear inequalities. In turn, these constraints define the shape of the 
fundamental diagram that is specified to the LWR model.

\begin{figure}[!ht]
  \centering
  \includegraphics[width=5.5cm,height=3.5cm]{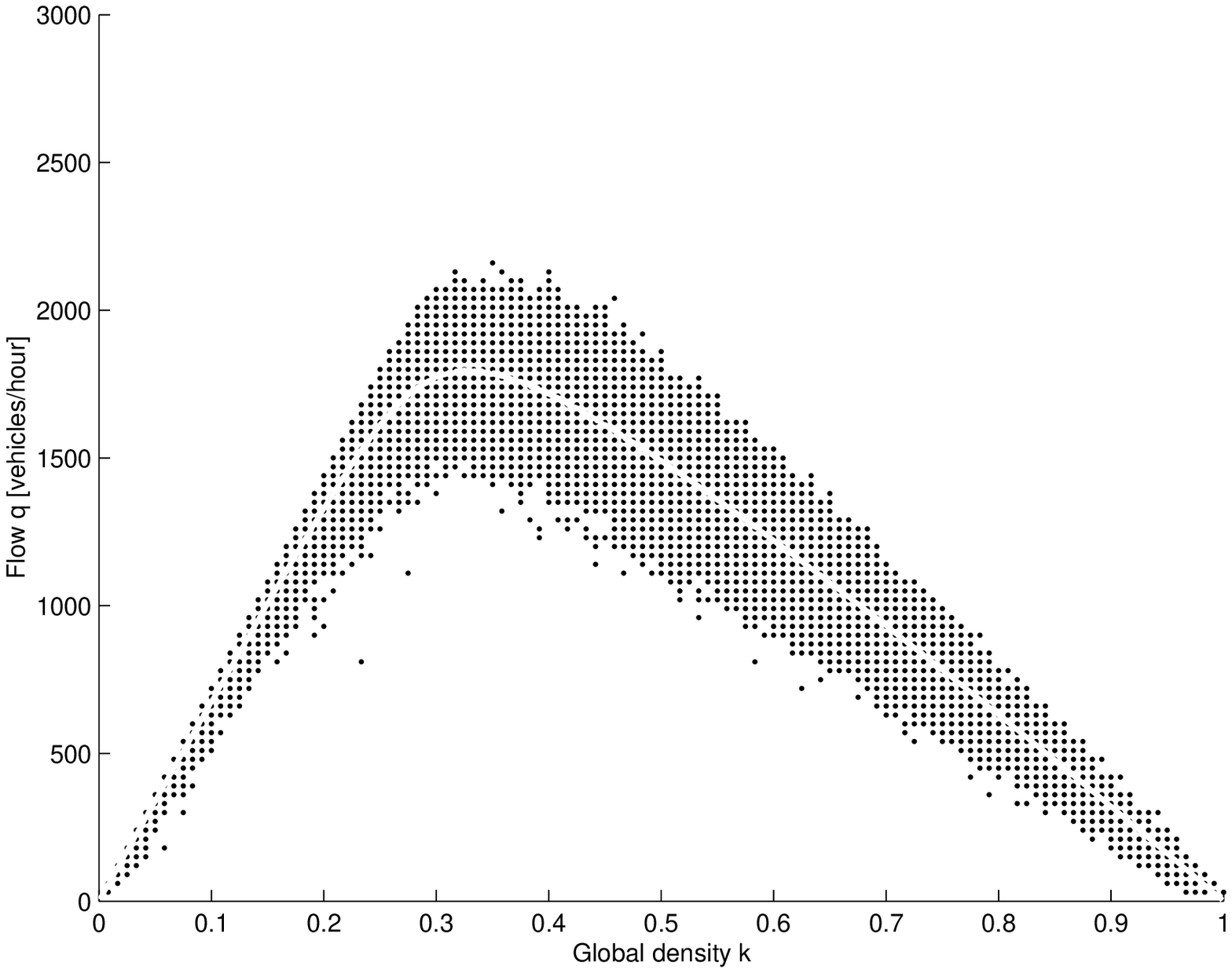}
  \includegraphics[width=5.5cm,height=3.5cm]{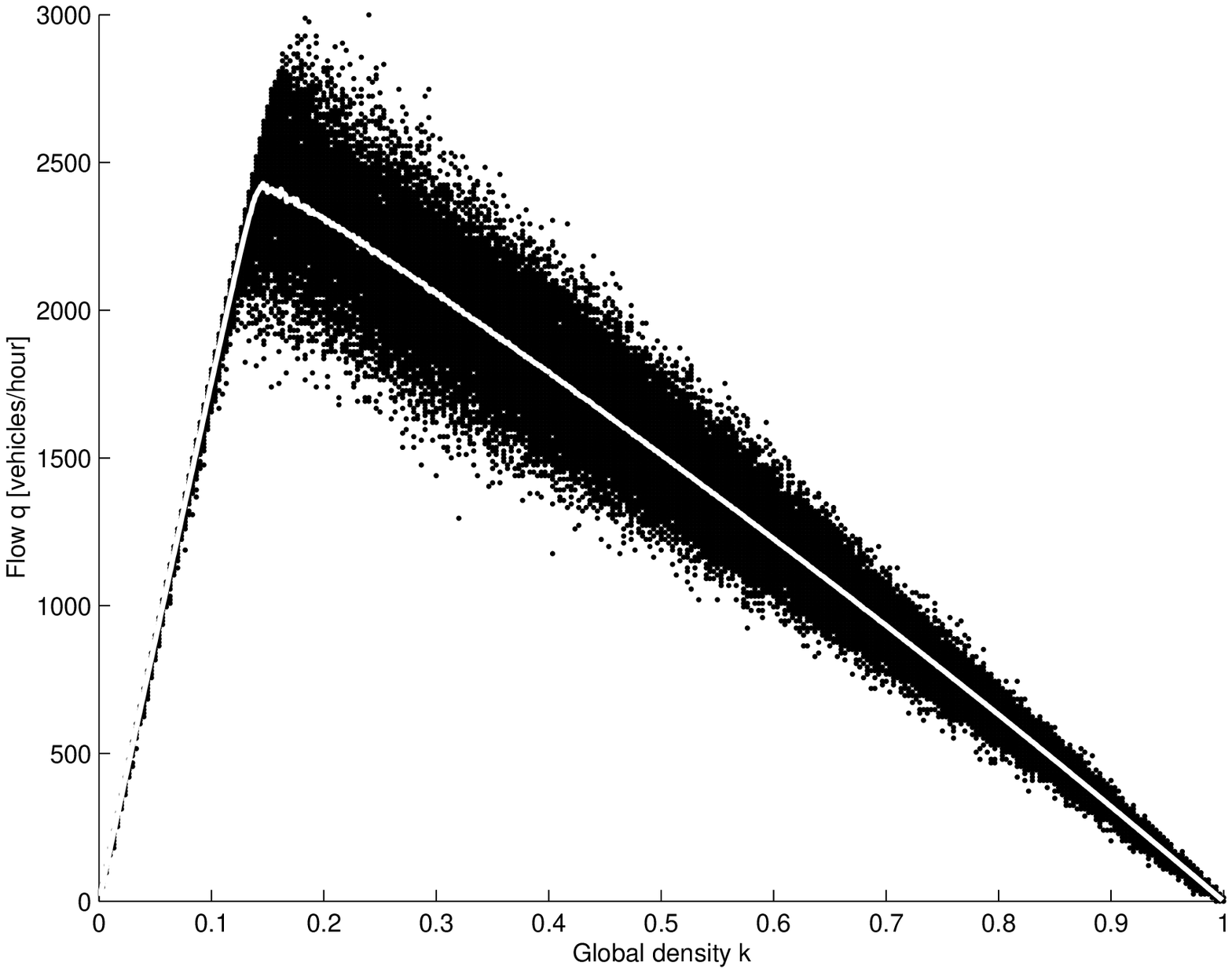}
  \caption{
    The ($k$,$q$) phase space diagrams of the STCA for $v_\text{max} = 
    2$~cells/s (left) and $v_\text{max} = 5$~cells/s (right). The stochastic 
    noise was $p = 0.1$ in both diagrams. The small points denote individual 
    measurements, whereas the white curves represent long-time averages.
  }
  \label{fig:STCAFDs}
\end{figure}

Our methodology sees the STCA complementary to the LWR model and vice versa, so 
the results can be of great assistance when interpreting the traffic dynamics in 
both models. Nevertheless, because the LWR model is only a coarse representation 
of reality, there are still some mismatches between the two approaches. One of 
the main concerns the authors discovered, is the fact that using a stationary 
fundamental diagram (i.e., an equilibrium relation between density and flow), 
always overestimates the practical capacity of a cellular automaton model (see 
e.g., figure \ref{fig:STCAFDs}, where the true capacity for $v_{\max} = 
5$~cells/s lies somewhere near 2400~vehicles/h, which is a rather low value).

Further research will focus on the dynamics of multi-lane traffic, on the 
heterogeneity of the traffic stream (using a heterogeneous LWR model), and on 
the relation between the capacity of a cellular automaton and the level of 
stochastic noise in the system.

% ================
% ACKNOWLEDGEMENTS
% ================
\section*{Acknowledgements}

Our research is supported by: \textbf{Research Council KUL}: GOA-Mefisto 666, 
GOA-AMBioRICS, several PhD/postdoc \& fellow grants, \textbf{FWO}: PhD/postdoc 
grants, projects, G.0240.99 (multilinear algebra), G.\-0407.\-02 (support vector 
machines), G.0197.02 (power islands), G.0141.03 (identification and 
cryptography), G.0491.03 (control for intensive care gly\-ce\-mia), G.0120.03 
(QIT), G.0452.04 (new quantum algorithms), G.0499.04 (robust SVM), research 
communities (ICCoS, ANMMM, MLDM), \textbf{AWI}: Bil. Int. Collaboration 
Hungary/Poland, \textbf{IWT}: PhD Grants, GBOU (McKnow), \textbf{Belgian Federal 
Science Policy Office}: IUAP P5/22 (`Dynamical Systems and Control: Computation, 
Identification and Modelling', 2002-2006), PODO-II (CP/40: TMS and 
Sustainability), \textbf{EU}: FP5-Quprodis, ERNSI, Eureka 2063-IMPACT, Eureka 
2419-FliTE, \textbf{Contract Research/agreements}: ISMC/IPCOS, Data4s, TML, 
Elia, LMS, Mastercard.

% ==========
% REFERENCES
% ==========
%\bibliographystyle{unsrt}
\bibliography{paper}

\end{document}